\documentclass[a4paper,num-refs]{perceiv-contemporary}

\journal{perceiv}

\usepackage{graphicx}
\usepackage{siunitx}
\usepackage{hyperref}
\usepackage{float}
\usepackage{amsmath}
\usepackage{booktabs}
\usepackage{multirow}
\usepackage{threeparttable}
\usepackage{caption}
\title{Prediction of cognitive decline for enrichment of Alzheimer’s disease clinical trials}

\author[1]{Angela Tam}
\author[1]{César Laurent}
\author[2,3]{Serge Gauthier}
\author[1,\authfn{1}]{Christian Dansereau}
\author[ \hspace{-1ex} \authfn{2}]{for the Alzheimer's Disease Neuroimaging Initiative}

\affil[1]{Perceiv Research Inc., Montréal, Québec, Canada}
\affil[2]{McGill University Research Centre for Studies in Aging, Montréal, Québec, Canada}
\affil[3]{Douglas Hospital Research Centre, McGill University, Montréal, Québec, Canada}

\authnote{\authfn{1}christian@perceiv.ai}
\authnote{\authfn{2}Data used in preparation of this article were obtained from the Alzheimer’s Disease Neuroimaging Initiative (ADNI) database (\url{adni.loni.usc.edu}). As such, the investigators within the ADNI contributed to the design and implementation of ADNI and/or provided data but did not participate in analysis or writing of this report. A complete listing of ADNI investigators can be found at: \url{http://adni.loni.usc.edu/wp-content/uploads/how_to_apply/ADNI_Acknowledgement_List.pdf}}

\runningauthor{Tam et al.}

\begin{document}

\begin{frontmatter}
\maketitle
\begin{abstract}
\textbf{Background:} A key issue to Alzheimer’s disease clinical trial failures is poor participant selection. Participants have heterogeneous cognitive trajectories and many do not decline during trials, which reduces a study’s power to detect treatment effects. Trials need enrichment strategies to enroll individuals who are more likely to decline.

\textbf{Objectives:} To develop machine learning models to predict cognitive trajectories in participants with early Alzheimer’s disease and presymptomatic individuals over 24 and 48 months respectively.

\textbf{Design:} Prognostic machine learning models were trained from a combination of demographics, cognitive tests, APOE genotype, and brain imaging data.

\textbf{Setting:} Data from the Alzheimer’s Disease Neuroimaging Initiative (ADNI), National Alzheimer’s Coordinating Center (NACC), Open Access Series of Imaging Studies (OASIS-3), PharmaCog, and a Phase 3 clinical trial in early Alzheimer’s disease were used for this study.

\textbf{Participants:} A total of 2098 participants who had demographics, cognitive tests, APOE genotype, and brain imaging data, as well as follow-up visits for 24-48 months were included.

\textbf{Measurements:} Baseline magnetic resonance imaging, cognitive tests, demographics, and APOE genotype were used to separate decliners, defined as individuals whose CDR - Sum of Boxes scores increased during a predefined time window, from stable individuals. A prognostic model to predict decline at 24 months in early Alzheimer’s disease was trained on 1151 individuals who had baseline diagnoses of mild cognitive impairment and Alzheimer’s dementia from ADNI and NACC. This model was validated on 115 individuals from a placebo arm of a Phase 3 clinical trial and 76 individuals from the PharmaCog dataset. A second prognostic model to predict decline at 48 months in presymptomatic populations was trained on 628 individuals from ADNI and NACC who were cognitively unimpaired at baseline. This model was validated on 128 individuals from OASIS-3.

\textbf{Results:} The models achieved up to 79\% area under the curve (cross-validated and out-of-sample). Power analyses showed that using prognostic models to recruit enriched cohorts of predicted decliners can reduce clinical trial sample sizes by as much as 51\% while maintaining the same detection power.

\textbf{Conclusions:} Prognostic tools for predicting cognitive decline and enriching clinical trials with participants at the highest risk of decline can improve trial quality, derisk endpoint failures, and accelerate therapeutic development in Alzheimer’s disease.

\end{abstract}

\begin{keywords}
Alzheimer's disease; clinical trials; cognitive decline; machine learning; trial enrichment
\end{keywords}
\end{frontmatter}

%%% Key points will be printed at top of second page
% \begin{keypoints*}
% \begin{itemize}
% \item This is the first point
% \item This is the second point
% \item One last point.
% \end{itemize}
% \end{keypoints*}

\pretolerance=8000
\tolerance=9000
\widowpenalty=10000
\clubpenalty=10000

%%%%%%%%%%%%%%%%%%%%%%%%%%%%%%%%%%%%%%%%
% INTRODUCTION
%%%%%%%%%%%%%%%%%%%%%%%%%%%%%%%%%%%%%%%%
\section{Introduction}
Over the past two decades, drug development in Alzheimer’s disease (AD) has been overwhelmingly disappointing \cite{Gauthier2016-dn,Aisen2020-oo}. The majority of AD clinical trials have failed to demonstrate statistically significant differences in cognitive trajectories between treatment and placebo groups \cite{Long2019-cy}. Failure to meet a cognitive endpoint may be partially due to individual heterogeneity in outcomes and poor participant selection.

Many late phase AD clinical trials have used a combination of clinical diagnostic criteria, thresholds on cognitive tests, genetic factors, and presence of amyloid pathology to select participants \cite{Honig2018-lh,Wessels2020-ej,Egan2019-pd,Budd_Haeberlein2020-zf,Swanson2021-rp,Ostrowitzki2017-iv,Mintun2021-wu,Burns2019-cw,Sperling2021-lq}. However, placebo groups in several randomized controlled trials for AD were reported to have highly variable rates of cognitive decline despite having similar inclusion/exclusion criteria \cite{Petersen2017-vh}. A study that modelled the effect of recruitment imbalance of decliners and stable individuals in simulated treatment and placebo groups suggested that any group differences reported in recent anti-amyloid trials could be partially explained by poor randomization, where there could have been either oversampling of decliners in the placebo arms or undersampling of decliners in the treatment arms \cite{Jutten2021-xq}. In fact, placebo groups often do not exhibit cognitive decline over the typical duration (18-24 months) of a trial \cite{Cummings2018-aj}, which poses a substantial challenge considering that a successful trial requires the placebo arm to experience a steeper cognitive decline than the treatment arm. Altogether, these findings suggest that current inclusion/exclusion criteria for AD trials are insufficient for recruiting participants with predictable cognitive trajectories.

Despite efforts to harmonize the definition of AD for research purposes \cite{Jack2018-hc}, AD is a heterogeneous disease with several distinct variants \cite{Lam2013-ju}. There is substantial heterogeneity across patients in their clinical presentations \cite{Scheltens2017-vq,Crane2017-nr}, and different subgroups of patients have different cognitive trajectories and distinct spatial patterns of brain atrophy and tau pathology \cite{Ossenkoppele2020-su}. Due to this heterogeneity, identifying individuals who are the most likely to experience cognitive decline during trials is the key to building high quality cohorts for successful trials. Enrichment strategies for enrolling likely decliners is especially important for preventative trials given that many individuals in earlier disease stages do not have detectable levels of pathology and that the presence of AD pathology has relatively low predictive accuracy of future decline in cognitively unimpaired individuals \cite{Dubois2021-ac}.

Many studies have focused on predicting diagnostic conversion (e.g. mild cognitive impairment (MCI) to dementia) (see Ansart et al 2021 \cite{Ansart2021-dv} for a review) as a potential enrichment strategy. However, a change in diagnosis can be a slow process that may not occur in the majority of participants over the duration of a clinical trial. Moreover, in trials with symptomatic participants, this strategy would restrain the selection to individuals with late-stage MCI who are on the cusp of converting to dementia and exclude potential candidates at earlier disease stages.

Encouraged by the recent successes of machine learning in the medical field \cite{Esteva2017dermatologist,Nam2019development}, we propose in the current study to improve the quality of AD clinical trials by using prognostic machine learning models to classify stable individuals and decliners as measured by change on a commonly used trial endpoint, the Clinical Dementia Rating Scale - Sum of Boxes (CDR-SB), within time windows that are typical of clinical trials \cite{Honig2018-lh,Egan2018-nq,Egan2019-pd,Wessels2020-ej,Swanson2021-rp,Ostrowitzki2017-iv,Budd_Haeberlein2020-zf,Mintun2021-wu,Lopez_Lopez2019-mw}. We trained these models using input features that are commonly collected during the screening process of clinical trials, i.e. magnetic resonance imaging (MRI), cognitive test scores, APOE $\epsilon$4 status, and demographics \cite{Honig2018-lh,Egan2018-nq,Egan2019-pd,Wessels2020-ej,Swanson2021-rp,Ostrowitzki2017-iv,Mintun2021-wu,Lopez_Lopez2019-mw,Sperling2020-pd}. With the objective of developing prognostic enrichment strategies across the spectrum of the disease continuum, we trained models for two populations: 1) individuals with early AD, i.e. participants with baseline diagnoses of AD dementia or MCI, and 2) presymptomatic individuals, i.e. participants who were cognitively unimpaired at baseline. The first model was trained on participants with early AD from both the Alzheimer’s Disease Neuroimaging Initiative (ADNI) and the National Alzheimer’s Coordinating Center (NACC) to predict decline over 24 months. It was then validated on two independent samples: a placebo arm of a Phase 3 clinical trial and the PharmaCog initiative. The second model was trained on presymptomatic individuals from both ADNI and NACC to predict decline over a longer period of 48 months. It was then validated on a separate cohort of individuals from the Open Access Series of Imaging Studies (OASIS-3). In both cases, we performed power analyses to quantify the quality of the enriched cohort of predicted decliners compared to the full unenriched cohort.

%%%%%%%%%%%%%%%%%%%%%%%%%%%%%%%%%%%%%%%
% METHODS
%%%%%%%%%%%%%%%%%%%%%%%%%%%%%%%%%%%%%%%
\section{Methods}

\subsection{Participants}
Participants from the Alzheimer’s Disease Neuroimaging Initiative (ADNI), National Alzheimer’s Coordinating Center (NACC), the third release from the Open Access Series of Imaging Studies (OASIS-3), PharmaCog, and the placebo group of an 18-month early AD Phase 3 clinical trial (NCT02080364) were included in this study. All participants gave informed consent to participate in these studies, which were approved by the research ethics committees of the institutions involved in data acquisition. Part of the data used in the preparation of this article was obtained from the Alzheimer’s Disease Neuroimaging Initiative (ADNI) database (adni.loni.usc.edu). The ADNI was launched in 2003 as a public-private partnership, led by Principal Investigator Michael W. Weiner, MD. The primary goal of ADNI has been to test whether serial magnetic resonance imaging (MRI), positron emission tomography (PET), other biological markers, and clinical and neuropsychological assessment can be combined to measure the progression of mild cognitive impairment (MCI) and early Alzheimer’s disease (AD). The NACC dataset comprised data for visits conducted between September 2005 and September 2020.

For the early AD discovery dataset, we included participants from ADNI and NACC who were at least 50 years of age and had a diagnosis of MCI or AD dementia at baseline. Inclusion/exclusion criteria for the ADNI dataset are described at \url{http://adni.loni.usc.edu/}. NACC is a multi-site dataset where each site has its own protocol for participant enrollment. For external validation, we used two datasets: 1) the placebo group of an 18-month Phase 3 clinical trial in participants with early AD, and 2) PharmaCog, a cohort of participants with baseline diagnoses of amnestic MCI. Key inclusion/exclusion criteria for the Phase 3 trial included a diagnosis of probable AD, Mini-Mental State Examination (MMSE) score between 21-26 and global CDR score of 0.5 or 1 at screening, and on a stable dose of cholinesterase inhibitors or memantine for at least 3 months prior to randomization. PharmaCog consisted of participants with amnestic MCI, and the inclusion/exclusion criteria for this dataset are described in Galluzi et al 2016 \cite{Galluzzi2016-ni}. Participants from each cohort were included if they had key demographics, APOE genotyping, cognitive test scores, MRI, and sufficient follow-up visits (24 months for discovery and PharmaCog, 18 months for the Phase 3 trial). 

For the presymptomatic discovery dataset, we included participants from ADNI and NACC who were at least 50 years of age and were cognitively unimpaired at baseline. For external validation, we selected participants from OASIS-3, a dataset provided by the Open Access Series of Imaging Studies, and its inclusion/exclusion criteria have been described in Lamontagne et al 2019 \cite{LaMontagne2019-fx}. Participants were included into our study if they were cognitively unimpaired at baseline, had key demographics, APOE genotyping, cognitive test scores, MRI, and at least 48 months of follow-up.

\subsection{Image acquisition and processing}
We analyzed 3D T1-weighted images acquired from 1.5 and 3.0 T MRI scanners (Siemens, Philips, and GE Medical Systems). For detailed descriptions of the MRI acquisition protocols, see \url{http://adni.loni.usc.edu/methods/documents/mri-protocols/} for ADNI, LaMontagne et al 2019 \cite{LaMontagne2019-fx} for OASIS-3, and Jovicich et al 2013 \cite{Jovicich2013-mx} for PharmaCog. This information is unavailable for NACC and the Phase 3 trial. 

Images were segmented into grey matter, white matter, and cerebrospinal fluid using Statistical Parametric Mapping (SPM) version 12 \cite{Friston2007-qn}. The volumes of the three tissue segmentations were summed together to calculate total intracranial volume (TIV) for each scan. The grey matter segmentations were normalized to MNI152 standard space and Jacobian scaled using the DARTEL toolbox \cite{Ashburner2007-ko} with a pre-defined group-specific template before being smoothed with an 8 mm isotropic full-width half maximum Gaussian kernel. From the normalized grey matter segmentations, we extracted features for all 170 regions in the AAL3 parcellation \cite{Rolls2020-qf} to use as inputs for our machine learning models.

\subsection{Outcome measure}
The Clinical Dementia Rating Scale - Sum of Boxes (CDR-SB) is a widely used cognitive outcome measure in AD clinical trials. The CDR-SB measures cognition and ability to function in 6 domains: memory, orientation, judgment and problem solving, community affairs, home and hobbies, and personal care. Generally, each domain is rated on a 5-point scale of levels of impairment: 0 for no impairment, 0.5 for “questionable”, 1 for “mild”, 2 for “moderate”, and 3 for “severe” \cite{Morris1993-pp}. The total CDR-SB score after summing the domain scores ranges from 0 to 18, where higher scores reflect worse impairment. We used change in CDR-SB score to define individuals as cognitively stable or decliners.

\subsection{Early AD: Task and model definition}
The first predictive task was to classify decliners and stable participants with MCI and AD dementia. Decliners were defined as individuals who had increased CDR-SB scores at 24 months of follow-up compared to baseline. Stable individuals were defined as participants who had no change or negative changes in CDR-SB scores at 24 months compared to baseline.

Baseline measures of age, sex, APOE $\epsilon$4 status (binarized as the presence of at least one $\epsilon$4 allele or not), cognitive test scores (CDR-SB, MMSE), total grey matter volume and total white matter volume normalized by TIV, and grey matter features extracted from AAL3 regions were used as inputs. The models used were Support Vector Machines (SVM) \cite{Cortes1995-cp}, preceded by a scaling to the [-1, 1] range, which were implemented in scikit-learn \cite{Pedregosa2011-uu}. Operating points were adjusted to obtain a specificity of 0.75 on the predicted stable individuals. Various combinations of demographic variables, clinical measures, and MRI were used as input features to generate multiple prediction models to assess which combination of features would yield the best performance. Hyper-parameters, including which feature set to use, were cross-validated in the discovery dataset using nested 5-fold cross-validation. The best performing model was further externally validated on the independent Phase 3 trial and PharmaCog datasets where the tasks were to identify decliners from stable individuals at the 18- and 24-month time points, respectively.

\subsection{Presymptomatic: Task and model definition}
The second predictive task was to classify decliners and stable individuals who were cognitively unimpaired at baseline. Decliners were defined as individuals who had increased CDR-SB scores at 48 months of follow-up compared to baseline. Stable individuals were defined as participants who had no change or negative changes in CDR-SB scores at 48 months compared to baseline.

Baseline measures of age, sex, APOE $\epsilon$4 status, years of education, cognitive test scores (CDR-SB, MMSE, Functional Activities Questionnaire), total grey matter volume and total white matter volume normalized by TIV, and grey matter features extracted from AAL3 regions were used as inputs. We used the exact same procedure described above for hyper-parameter tuning. The best model was further externally validated on \mbox{OASIS-3.}

\subsection{Power analysis and sample size estimates}
We followed Noordzij et al 2010 \cite{Noordzij2010-zj} to estimate the minimum sample size necessary to detect a reduction in average rates of cognitive decline between the treatment and placebo arms. This method assumes linear rates of decline, a crude but reasonable assumption when considering a single follow-up time point. We used a two-sided test with a significance level $\alpha=0.05$ and power of 80\% ($\beta=0.2$). The required sample size $n$ can then be calculated as follows:
\begin{align}
    n = 2\sigma^2\left(\frac{a+b}{\mu_t - \mu_p}\right)^2
\end{align}
where $\mu_t$ and $\mu_p$ are the population averages in the treatment and placebo arms, respectively, $\sigma$ is the standard deviation of the overall population, and $a$ $(=1.96)$ and $b$ $(=0.842)$ are the multipliers for $\alpha$ and $\beta$, respectively \cite{Noordzij2010-zj}. In our analyses, the treatment effect, which is the difference in population averages $\mu_t - \mu_p$ in Equation 1, was set to 0.3, which meant we wanted to observe at least a 30\% difference in cognitive decline between the treatment and placebo arms. A 30\% treatment effect was chosen based on past clinical trials that assumed drug effect sizes between 25\% and 35\% for their power calculations \cite{Egan2018-nq, Egan2019-pd, Mintun2021-wu, Wessels2020-ej}. Equation 1 can thus be used to separately compute sample sizes for both the unenriched and enriched cohorts, and the relative difference between these two quantities can be used to assess the quality of the enriched cohort with respect to the full unenriched cohort.

\section{Results}
%%%%%%%%%%%%%%%%%%%%%%%%%%%%%%%%%%%%%%%%
% RESULTS: EARLY AD
%%%%%%%%%%%%%%%%%%%%%%%%%%%%%%%%%%%%%%%%

\subsection{Prediction of cognitive decline in early AD}

\paragraph{Early AD: Participants}
A discovery dataset of 1151 participants with early AD was pooled from ADNI ($n=858$) and NACC ($n=293$). 115 individuals from the Phase 3 clinical trial and 76 individuals from the PharmaCog dataset were included as two independent validation cohorts. Participant characteristics for each of the individual datasets are described in supplementary \autoref{app:t1}. 

In the aim of curating a large and diverse discovery dataset, we decided against using inclusion/exclusion criteria based on thresholds on cognitive scores or amyloid positivity. However, the majority of the early AD discovery participants had baseline scores of MMSE and global CDR that were within the ranges of inclusion/exclusion criteria (e.g. MMSE $\geq$ 20, global CDR of $0.5-1$) of previous clinical trials in early AD \cite{Honig2018-lh,Egan2019-pd,Wessels2020-ej,Swanson2021-rp,Ostrowitzki2017-iv,Mintun2021-wu}. See \autoref{app:f1} in supplementary material for details about the distributions of the variables typically used for inclusion/exclusion criteria for all of the aforementioned datasets.

%%% FIGURE 1: EARLY AD
\begin{figure*}[ht]
    \centering
        \includegraphics[width=\linewidth]{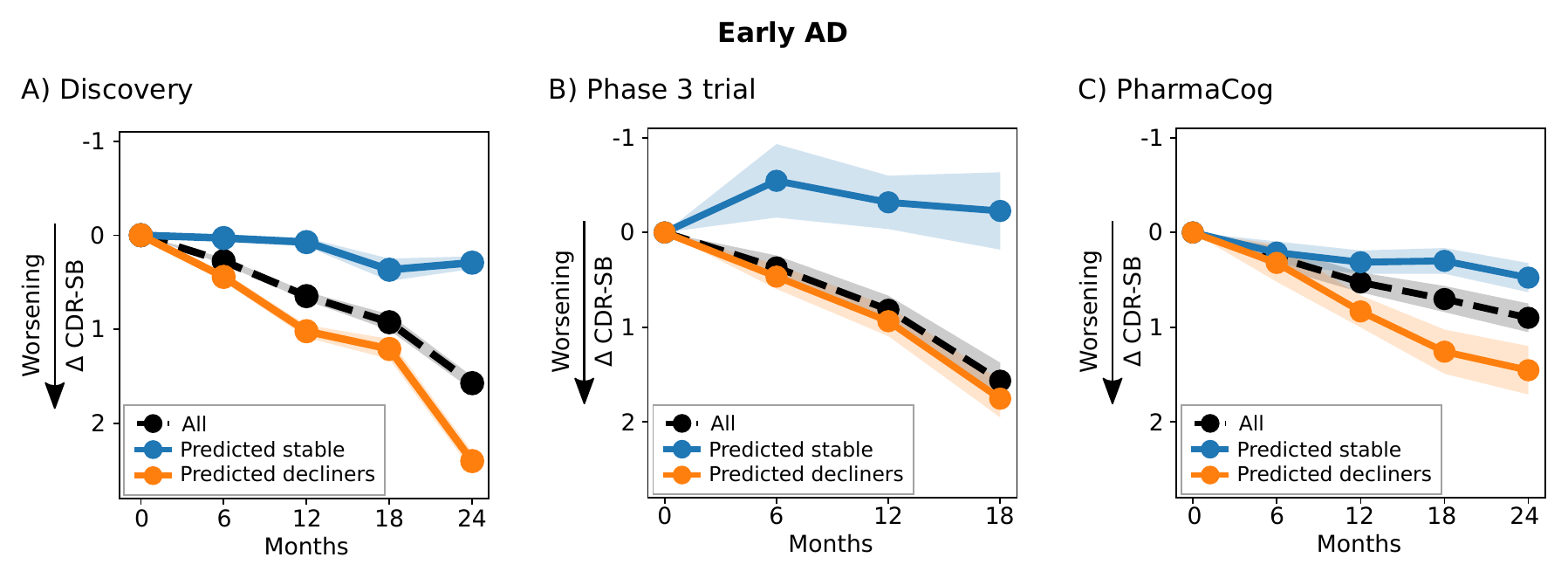}
        \caption{Stratification of predicted cognitive decliners and stable individuals in early AD participants. Plots illustrate the mean change in cognition measured by the CDR-SB over time in the predicted stable (solid blue line), predicted decliners (solid orange line), and the entire cohort (dashed black line) in A) the discovery dataset, B) Phase 3 trial validation cohort, and C) PharmaCog validation cohort. The shaded areas represent the standard errors of the means. CDR-SB: Clinical Dementia Rating Scale - Sum of boxes.}
        \label{fig:f1}
\end{figure*}

%%% TABLE 1: EARLY AD
\begin{table*}[ht]
\caption{Characteristics of the early AD discovery and external validation cohorts}\label{tab:t1}
\begin{tabularx}{\linewidth}{ 
@{}>{\smaller\raggedright}p{2.75cm} 
@{}>{\smaller\centering}p{1.5cm}
@{}>{\smaller\centering}p{1.5cm}
@{}>{\smaller\centering}p{1.5cm}
@{}>{\smaller\raggedright}p{0.45cm}
@{}>{\smaller\centering}p{1.5cm}
@{}>{\smaller\centering}p{1.5cm}
@{}>{\smaller\centering}p{1.5cm}
@{}>{\smaller\raggedright}p{0.45cm}
@{}>{\smaller\centering}p{1.5cm}
@{}>{\smaller\centering}p{1.5cm}
@{}>{\smaller\centering\arraybackslash}p{1.5cm}@{} }
\toprule
 & \multicolumn{3}{c}{\bf Early AD discovery} &  & \multicolumn{3}{c}{\bf Phase 3 trial validation} & &  \multicolumn{3}{c}{\bf PharmaCog validation}\\
\cmidrule{2-4} \cmidrule{6-8} \cmidrule{10-12}
 & \bf All & \bf Predicted stable & \bf Predicted decliners & & \bf All & \bf Predicted stable & \bf Predicted decliners & & \bf All & \bf Predicted stable & \bf Predicted decliners\\
\midrule
N                                                   & 1151       & 452        & 699                              & & 115        & 11          & 104                             & & 76         & 43         & 33\\
Baseline age                                        & 73.1 (7.9) & 71.1 (7.8) & 74.5 (7.6)\textsuperscript{*}    & & 74.3 (7.9) & 71.4 (6.8)  & 74.6 (8.0)                      & & 69.2 (7.5) & 67.2 (6.7) & 72.0 (7.6)\textsuperscript{*}\\
Female \%                                           & 43.5       & 43.6       & 43.5                             & & 47.0       & 54.5        & 46.1                            & & 55.5       & 55.8       & 54.5\\
APOE $\epsilon$4 carriers \%                        & 53.3       & 33.0       & 66.5\textsuperscript{*}          & & 51.3       & 18.2        & 54.8\textsuperscript{*}         & & 47.4       & 34.9       & 63.6\textsuperscript{*}\\
Education                                           & 15.6 (3.0) & 16.0 (2.8) & 15.4 (3.1)\textsuperscript{*}    & & -          & -           & -                               & & 11.5 (4.5) & 11.2 (4.6) & 11.9 (4.3)\\
Amyloid positive \%\textsuperscript{a}              & 67.9       & 48.9       & 84.7\textsuperscript{*}          & & -          & -           & -                               & & 42.1       & 34.9       & 51.5\\
Baseline MMSE                                       & 26.3 (3.3) & 28.2 (1.6) & 25.0 (3.5)\textsuperscript{*}    & & 23.4 (1.7) & 24.1 (1.0)  & 23.4 (1.7)                      & & 26.7 (1.8) & 27.3 (1.7) & 25.9 (1.7)\textsuperscript{*}\\
Baseline CDR-SB                                     & 2.31 (1.9) & 1.31 (0.8) & 2.96 (2.1)\textsuperscript{*}    & & 3.99 (1.6) & 3.73 (1.7)  & 4.01 (1.6)                      & & 1.10 (0.8) & 0.93 (0.7) & 1.33 (1.0)\\
Change in CDR-SB at endpoint\textsuperscript{b}     & 1.57 (2.4) & 0.29 (1.4) & 2.40 (2.6)\textsuperscript{*}    & & 1.56 (2.0) & -0.23 (1.4) & 1.75 (2.0)\textsuperscript{*}   & & 0.90 (1.3) & 0.48 (1.0) & 1.45 (1.5)\textsuperscript{*}\\
True decliners \%                                   & 64.6       & 36.5       & 82.7\textsuperscript{*}          & & 72.2       & 36.4        & 76.0\textsuperscript{*}         & & 53.9       & 39.5       & 72.7\textsuperscript{*}\\
\bottomrule
\end{tabularx}
\begin{tablenotes}[flushleft]
\smaller
\item All values are means (standard deviations) unless otherwise stated.
\item Abbreviations: APOE, apolipoprotein E; CDR-SB, Clinical Dementia Rating Scale - Sum of boxes; MMSE, Mini-Mental State Examination.
\item \textsuperscript{a}Only 669 individuals in the discovery cohort and 76 individuals in PharmaCog had documented amyloid status via either positron emission tomography (PET) or cerebrospinal fluid (CSF).
\item \textsuperscript{b}Endpoints were 24 months for the discovery and PharmaCog validation datasets and 18 months for the Phase 3 trial validation dataset.
\item \textsuperscript{*}The predicted stable and predicted decliners are significantly different within each cohort ($p < 0.05$, two-sided) as assessed by Mann-Whitney U tests for continuous variables and chi-squared tests for categorical variables.
\end{tablenotes}
\end{table*}

\paragraph{Early AD: Model}
The model to predict decline at 24 months in participants with early AD that obtained the highest internal validation area under the curve (AUC) used baseline measures of age, sex, APOE $\epsilon$4 carriership, MMSE, CDR-SB, and volumes from anatomical brain regions extracted from MRI as input features. See supplementary \autoref{app:t2} for the internal validation performances of models using different sets of features. The best model was further externally validated, and its performance is presented in the main results (\autoref{fig:f1}, \autoref{tab:t1}). Gray matter volumes of regions in the temporal lobe, baseline MMSE score, and APOE $\epsilon$4 status comprised the top ranking features that contributed to the model output (see supplementary \autoref{app:f2} for details).

\paragraph{Early AD: Stratification and power analysis}
In the discovery dataset, the best model achieved a mean ($\pm$ std) test AUC of $78.7 \pm 4.9\%$, sensitivity of $77.8 \pm 4.5\%$, specificity of $70.4 \pm 5.9\%$, and precision of $82.7 \pm 3.0\%$. The whole discovery sample had a mean ($\pm$ std) change of 1.57 $\pm$ 2.4 points on the CDR-SB at 24 months compared to baseline. 408 individuals actually remained stable over 24 months. From baseline data only, the predictive model classified 452 individuals as stable and 699 as decliners. The predicted stable group had a mean change of $0.29 \pm 1.4$ points on the CDR-SB at 24 months compared to baseline. On the other hand, the predicted decliners had a mean change of $2.40 \pm 2.6$ points (\autoref{fig:f1}A, \autoref{tab:t1}).

On average, the predicted stable individuals were younger, more educated, and had lower CDR-SB scores and higher MMSE scores at baseline than the predicted decliners (\autoref{tab:t1}). On the other hand, the predicted decliners were more likely to be APOE $\epsilon$4 carriers and amyloid positive. However, it is important to note that the distributions of each of these variables (supplementary \autoref{app:f1}) overlap almost completely between the predicted classes, which shows that the class separation would not be possible by relying only on tighter inclusion/exclusion criteria. Furthermore, there is also an almost complete overlap between these distributions and the ones of the full discovery sample, which indicates the model separated stable individuals from decliners across the whole spectrum of age and cognitive ability. 

A power analysis was performed to show that, by using the full unenriched discovery sample, a two-arm clinical trial would require 422 individuals per arm in order to detect a 30\% difference in CDR-SB change at 80\% power. On the other hand, by selectively using the enriched cohort of predicted decliners, only 206 individuals per arm would be required for enrollment. Recruiting specifically for likely decliners would thus allow for a 51.2\% reduction in sample size compared to recruiting the full unenriched discovery sample, while still maintaining the same study power.

\paragraph{Early AD: External validation}
In an independent validation cohort derived from the placebo arm of a Phase 3 clinical trial in early AD, the model obtained 71.4\% AUC, 95.2\% sensitivity, 21.9\% specificity, and 76.0\% precision for predicting decliners at 18 months. 32 individuals actually remained stable throughout the 18 months and the whole sample had a mean change of $1.56 \pm 2.0$ points on the CDR-SB at the endpoint compared to baseline. As a group, the 11 predicted stable individuals improved on the CDR-SB at 18 months with a mean change of $-0.23 \pm 1.4$ points, whereas the 104 predicted decliners had a mean change of $1.75 \pm 2.0$ (\autoref{fig:f1}B, \autoref{tab:t1}).

In a second validation cohort, PharmaCog, the model obtained 72.2\% AUC, 58.5\% sensitivity, 74.3\% specificity, and 72.7\% precision for predicting decliners at 24 months. 35 individuals actually remained stable across 24 months. The whole sample experienced a mean change of $0.90 \pm 1.3$ points on the CDR-SB at 24 months compared to baseline. The 43 predicted stable individuals had a mean change of $0.48 \pm 1.0$ points on the CDR-SB, while the 33 predicted decliners had a mean change of $1.45 \pm 1.5$ (\autoref{fig:f1}C, \autoref{tab:t1}).

\begin{figure}[h!]
    \centering
        \includegraphics[width=0.7\linewidth]{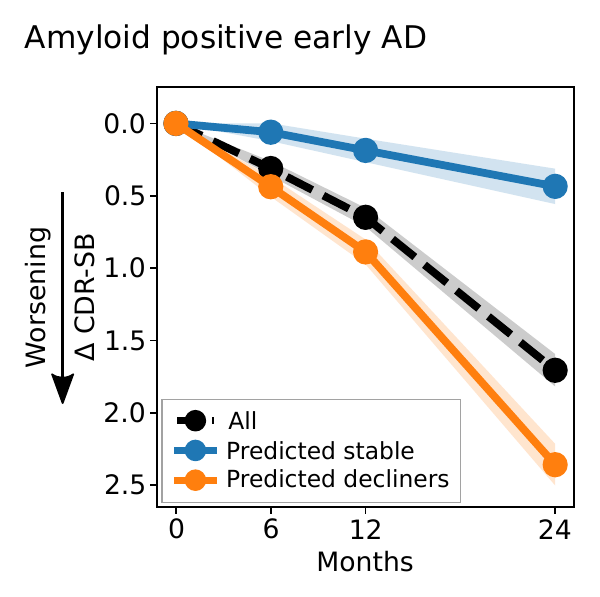}
        \caption{Stratification of predicted cognitive decliners and stable individuals in amyloid positive early AD participants in the discovery dataset ($n=454$). The plot illustrates the mean change in cognition measured by the CDR-SB over time in the predicted stable (solid blue line), predicted decliners (solid orange line), and the entire cohort (dashed black line). The shaded areas represent the standard errors of the means. CDR-SB: Clinical Dementia Rating Scale - Sum of boxes.}
        \label{fig:f2}
\end{figure}

%%%%%%%%%%%%%%%%%%%%%%%%%%%%%%%%%%%%%%%%
% RESULTS: EARLY AD AMYLOID POSITIVE
%%%%%%%%%%%%%%%%%%%%%%%%%%%%%%%%%%%%%%%%
\paragraph{Early AD: Stratification of amyloid positive individuals}
A subset of amyloid positive individuals in the discovery dataset ($n=454$) was further examined. Similarly to the entire cohort, there was a fraction (31.1\%) of amyloid positive individuals who did not exhibit decline on the CDR-SB at 24 months of follow-up (\autoref{tab:t2}). Of these 454 individuals, the model classified 154 as stable and 300 as decliners and achieved $78.6 \pm 7.0\%$ AUC, $80.5 \pm 4.6\%$ sensitivity, $66.6 \pm 6.4\%$ specificity, and $84.0 \pm 4.6\%$ precision. The predicted stable group had a mean change of $0.43 \pm 1.5$ points on the CDR-SB at 24 months, while the predicted decliners had a mean change of $2.36 \pm 2.5$ (\autoref{fig:f2}, \autoref{tab:t2}).

A power analysis was performed to show that, when using an unenriched cohort of amyloid positive individuals, a two-arm clinical trial would require 343 participants per arm to detect a 30\% treatment effect at 80\% power. By using the enriched cohort of amyloid positive decliners, only 195 participants per arm would be required, yielding a sample size reduction of 43.1\% compared to the original unenriched cohort.

%%%%%%%%%%%%%%%%%%%%%%%%%%%%%%%%%%%%%%%
% RESULTS: PRESYMPTOMATIC
%%%%%%%%%%%%%%%%%%%%%%%%%%%%%%%%%%%%%%%

\subsection{Prediction of cognitive decline in presymptomatic individuals}

\paragraph{Presymptomatic: Participants}
The discovery cohort included 628 participants who were pooled from ADNI ($n=237$) and NACC ($n=391$). The validation cohort was a sample of 128 participants from OASIS-3. Participant characteristics for both cohorts are detailed in supplementary \autoref{app:t1}. Similarly to the early AD discovery dataset, we did not select any presymptomatic participants based on thresholds on cognitive scores or amyloid positivity in order to increase the size and diversity of our samples. The majority of the presymptomatic participants had baseline scores of MMSE greater than 24 and global CDR of 0 (supplementary \autoref{app:f1}), which broadly fit the cognitive inclusion/exclusion criteria of past and ongoing preventative trials for AD \cite{Burns2021-zz,Sperling2021-lq,Sperling2020-pd,Szabo-Reed2019-vn}.

%%% FIGURE 2: EARLY AD amyloid positive

%%% TABLE 2: EARLY AD amyloid positive
\begin{table}[h!]
\caption{Characteristics of the amyloid positive individuals with early AD}\label{tab:t2}
\begin{tabularx}{\linewidth}{
@{}>{\smaller\raggedright}p{3.85cm} 
@{}>{\smaller\centering}p{1.5cm}
@{}>{\smaller\centering}p{1.5cm}
@{}>{\smaller\centering\arraybackslash}p{1.5cm}@{} }
\toprule
                                & \bf All       & \bf Predicted stable      & \bf Predicted decliners\\
\midrule
N                               & 454           & 154           & 300\\
Baseline age                    & 73.4 (7.4)    & 71.7 (7.5)    & 74.3 (7.2)\textsuperscript{*}\\
Female \%                       & 41.8          & 41.6          & 42.0\\
APOE $\epsilon$4 carriers \%    & 66.9          & 50.0          & 75.7\textsuperscript{*}\\
Education                       & 15.8 (2.9)    & 16.2 (2.6)    & 15.6 (3.0)\textsuperscript{*}\\
Baseline MMSE                   & 26.5 (2.6)    & 28.2 (1.6)    & 25.6 (2.5)\textsuperscript{*}\\
Baseline CDR-SB                 & 2.18 (1.6)    & 1.32 (0.8)    & 2.62 (1.7)\textsuperscript{*}\\
Change in CDR-SB at 24 months   & 1.71 (2.4)    & 0.43 (1.5)    & 2.36 (2.5)\textsuperscript{*}\\
True decliners \%               & 68.9          & 39.6          & 84.0\textsuperscript{*}\\
\bottomrule
\end{tabularx}
\begin{tablenotes}[flushleft]
\smaller
\item All values are means (standard deviations) unless otherwise stated.
\item Abbreviations: APOE, apolipoprotein E; CDR-SB, Clinical Dementia Rating Scale - Sum of boxes; MMSE, Mini-Mental State Examination.
\item \textsuperscript{*}The predicted stable and predicted decliners are significantly different ($p < 0.05$, two-sided) as assessed by Mann-Whitney U tests for continuous variables and chi-squared tests for categorical variables.
\end{tablenotes}
\end{table}

\begin{figure*}[t!] 
    \centering
        \includegraphics[width=0.7\linewidth]{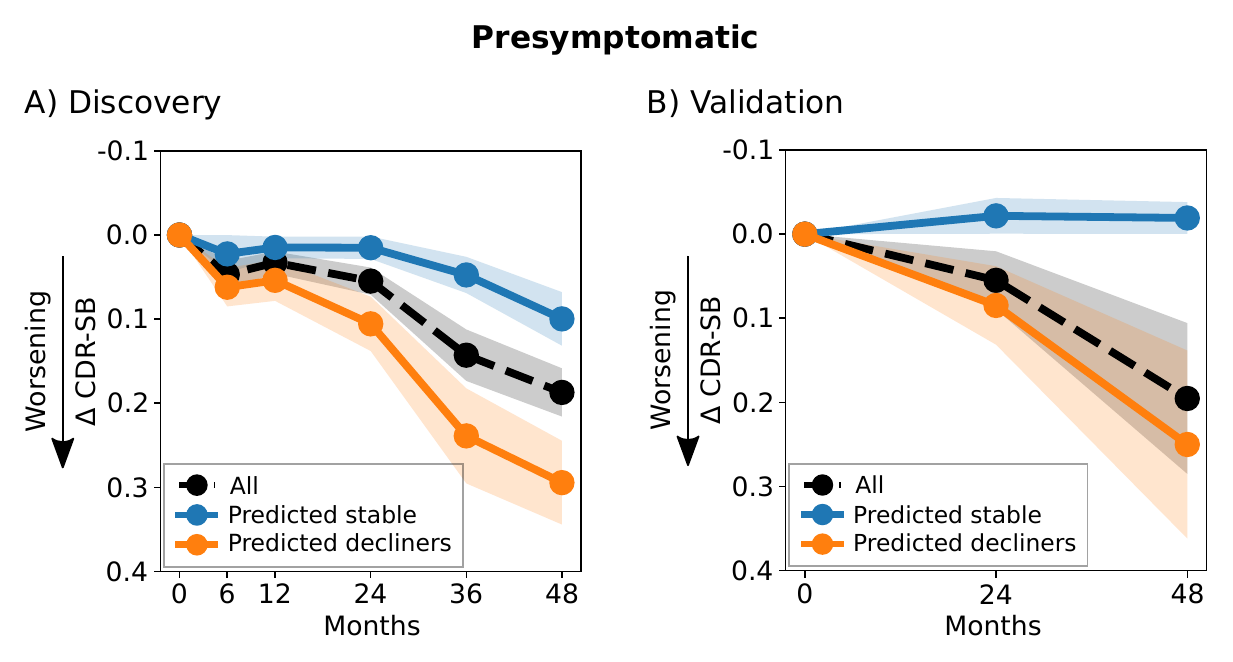}
        \caption{Stratification of predicted cognitive decliners and stable individuals in presymptomatic participants. Plots illustrate the mean change in cognition measured by the CDR-SB over time in the predicted stable (solid blue line), predicted decliners (solid orange line), and the entire cohort (dashed black line) in A) the discovery dataset, and B) the validation cohort. The shaded areas represent the standard errors of the means. CDR-SB: Clinical Dementia Rating Scale - Sum of boxes.}
        \label{fig:f3}
\end{figure*}

\paragraph{Presymptomatic: Model}
The best model to predict decline at 48 months in presymptomatic participants used baseline measures of age, sex, APOE $\epsilon$4 carriership, education, MMSE, CDR-SB, Functional Activities Questionnaire, and volumes from anatomical brain regions extracted from MRI as input features. Internal validation performances of models using different sets of features are reported in supplementary \autoref{app:t3}. The best model was further externally validated and is presented in \autoref{fig:f3} and \autoref{tab:t3}. Baseline age, APOE $\epsilon$4 status, and regional deep gray matter volumes were among the top ranking features that contributed to the model output (see supplementary \autoref{app:f2}).

\paragraph{Presymptomatic: Stratification and power analysis}
In the discovery dataset, the best model achieved a mean test AUC of $71.7 \pm 4.2\%$, sensitivity of $70.0 \pm 7.7\%$, specificity of $60.3 \pm 3.4\%$, and precision of $26.6 \pm 2.4\%$. The discovery cohort had a mean change of $0.19 \pm 0.7$ points on the CDR-SB at 48 months compared to baseline. 521 individuals in the sample did not actually decline over 48 months. The predictive model classified 346 participants as stable and 282 as decliners. The predicted stable group had a mean change of $0.10 \pm 0.6$ points on the CDR-SB at 48 months compared to baseline. On the other hand, the predicted decliners had a mean change of $0.29 \pm 0.8$ (\autoref{fig:f3}A, \autoref{tab:t3}).

The predicted stable group was younger, had a lower proportion of APOE $\epsilon$4 carriers, and had higher MMSE scores at baseline than the predicted decliners (\autoref{tab:t3}). However, as in our early AD experiments, there was an almost complete overlap of the distributions, not only between the two classes, but also with the full cohort (supplementary \autoref{app:f1}).

A power analysis was performed to show that, using the full unenriched discovery sample, a clinical trial would require 2577 individuals per arm in order to detect a 30\% difference in CDR-SB change at 80\% power. However, using the enriched cohort of predicted decliners, only 1412 individuals would be required for enrollment, which amounts to a 45.2\% sample size reduction from recruiting the unenriched cohort.

%%% FIGURE 3: PRESYMPTOMATIC

%%% TABLE 3: PRESYMPTOMATIC
\begin{table*}[t!]
\centering
\begin{threeparttable}
\caption{Characteristics of the presymptomatic discovery and validation cohorts}
\label{tab:t3}
\begin{tabularx}{0.78\linewidth}{ 
@{}>{\smaller\raggedright}p{3.9cm} 
@{}>{\smaller\centering}p{1.5cm}
@{}>{\smaller\centering}p{1.5cm}
@{}>{\smaller\centering}p{1.5cm}
@{}>{\smaller\raggedright}p{0.45cm}
@{}>{\smaller\centering}p{1.5cm}
@{}>{\smaller\centering}p{1.5cm}
@{}>{\smaller\centering\arraybackslash}p{1.5cm}@{} }
\toprule
 & \multicolumn{3}{c}{\bf Presymptomatic discovery} &   & \multicolumn{3}{c}{\bf Validation}\\
\cmidrule{2-4} \cmidrule{6-8}
                & \bf All & \bf Predicted stable & \bf Predicted decliners &            &   \bf All & \bf Predicted stable & \bf Predicted decliners\\
\midrule
N                                           & 628           & 346           & 282                               & & 128         & 26            & 102\\
Baseline age                                & 69.3 (9.0)    & 64.7 (7.8)    & 74.9 (7.0)\textsuperscript{*}     & & 69.4 (5.4)  & 67.5 (2.6)    & 69.9 (5.8)\textsuperscript{*}\\
Female \%                                   & 62.1          & 64.7          & 58.9                              & & 53.1        & 34.6          & 57.8\\
APOE $\epsilon$4 carriers \%                & 30.9          & 30.6          & 31.2                              & & 34.4        & 19.2          & 38.2\\
Education                                   & 16.3 (2.8)    & 16.5 (2.5)    & 16.0 (3.0)                        & & 16.0 (2.4)  & 16.1 (2.0)    & 16.0 (2.5)\\
Amyloid positive \%\textsuperscript{a}      & 36.1          & 26.7          & 42.2\textsuperscript{*}           & & 29.0        & 14.3          & 32.7\\
Baseline MMSE                               & 29.1 (1.1)    & 29.4 (0.9)    & 28.9 (1.3)\textsuperscript{*}     & & 29.0 (1.5)  & 29.1 (0.9)    & 28.8 (1.6)\\
Baseline CDR-SB                             & 0.07 (0.3)    & 0.04 (0.1)    & 0.11 (0.4)                        & & 0.02 (0.1)  & 0.02 (0.1)    & 0.02 (0.1)\\
Change in CDR-SB at 48 months               & 0.19 (0.7)    & 0.10 (0.6)    & 0.29 (0.8)\textsuperscript{*}     & & 0.19 (1.0)  & -0.02 (0.1)   & 0.25 (1.1)\\
True decliners \%                           & 17.0          & 9.2           & 26.6\textsuperscript{*}           & & 10.2        & 0.0           & 12.7\\
\bottomrule
\end{tabularx}
\begin{tablenotes}[flushleft]
\smaller
\item All values are means (standard deviations) unless otherwise stated.
\item Abbreviations: APOE, apolipoprotein E; CDR-SB, Clinical Dementia Rating Scale - Sum of boxes; MMSE, Mini-Mental State Examination.
\item \textsuperscript{a}Only subsets of the discovery ($n=191$) and validation ($n=69$) datasets had documented amyloid status via either positron emission tomography (PET) or cerebrospinal fluid (CSF).
\item \textsuperscript{*}The predicted stable and predicted decliners are significantly different within each dataset ($p < 0.05$, two-sided) as assessed by Mann-Whitney U tests for continuous variables and chi-squared tests for categorical variables.
\end{tablenotes}
\end{threeparttable}
\end{table*}

\paragraph{Presymptomatic: External validation}
In the independent validation cohort, the model obtained 77.6\% AUC, 100.0\% sensitivity, 22.6\% specificity, and 12.7\% precision. 115 individuals actually remained stable throughout 48 months, and the entire sample had a mean change of $0.19 \pm 1.0$ points on the CDR-SB at 48 months compared to baseline. The 26 predicted stable individuals had a mean change of $-0.02 \pm 0.1$ points on the CDR-SB at 48 months, while the 102 predicted decliners had a mean change of $0.25 \pm 1.1$ (\autoref{fig:f3}B, \autoref{tab:t3}).

%%%%%%%%%%%%%%%%%%%%%%%%%%%%%%%%%%%%%%%
% DISCUSSION
%%%%%%%%%%%%%%%%%%%%%%%%%%%%%%%%%%%%%%%
\section{Discussion}
We developed machine learning models to separate individuals who are likely to experience cognitive decline from those who will remain stable over periods of 24 and 48 months in participants with early AD and presymptomatic populations, respectively. We trained our models on widely accessible measures (demographics, APOE $\epsilon$4 status, cognitive test scores, MRI), which are typically collected during trial recruitment phases, to detect future decline, measured by change in CDR-SB, a commonly used endpoint in AD clinical trials. We validated these models on independent datasets and demonstrated they can also enrich for decliners in amyloid positive cohorts. Finally, we ran power analyses to show that, by specifically selecting likely decliners, sample sizes can be reduced by as much as 51\% compared to enrolling all eligible participants from an unenriched population, while still retaining the same statistical power.

It is interesting to note that even among amyloid positive participants with early AD, a third did not exhibit decline on the CDR-SB over the course of two years (\autoref{tab:t2}), which demonstrates that amyloid positivity alone does not guarantee cognitive decline. This suggests that, despite movements encouraging the use of biomarkers for trial enrichment \cite{Wolz2016-sp,Mattsson2015-hg}, exclusively selecting amyloid positive participants is an insufficient enrichment strategy to meet clinical endpoints. Our models were nonetheless able to stratify amyloid positive decliners from stable individuals with similar accuracy as in the full discovery cohort of early AD participants. Unfortunately, small sample sizes of amyloid positive individuals (Phase 3 trial $n=n/a$, PharmaCog $n=33$, presymptomatic discovery $n=69$, presymptomatic validation $n=20$) precluded us from replicating this finding in other datasets. Our results show that targeted selection of decliners can further enrich amyloid positive cohorts with individuals who will experience cognitive decline.

Previous works that modelled disease trajectories in AD have mainly focused on predicting diagnostic changes, where prognostic models that classify MCI patients who progress to AD dementia vs stable individuals are quite prominent (see Ansart et al 2021 \cite{Ansart2021-dv} for a review). However, the prediction of risk of conversion to dementia is a suboptimal strategy for trial enrichment as only 9.8\% to 36.3\% of MCI patients progress to AD dementia within two years of follow-up \cite{Ward2013-uu}. Models that are more relevant to clinical trial enrichment would be predicting the quantitative change in cognitive scores \cite{Liem2020-gf}, forecasting outcomes and biomarker evolution \cite{Marinescu2019-bc}, or predicting decliners and non-decliners on common cognitive endpoints \cite{Bhagwat2018-ys,Shafiee2021-mr} as in the current study.

The results from our analysis of feature importance (Supplementary \autoref{app:f2}) support previous literature that have shown certain factors, such as APOE $\epsilon$4 status and brain atrophy \cite{Reitz2011-qm}, are associated with a higher risk of future cognitive decline. However, while it may be tempting to establish binary inclusion/exclusion criteria on these factors to recruit trial participants, it is worth noting such a strategy is suboptimal. For instance, although APOE $\epsilon$4 status was a top ranking feature in our models, approximately 40\% and 60\% of true decliners in the early AD and presymptomatic groups, respectively, were APOE $\epsilon$4 negative. Thus, recruiting specifically for APOE $\epsilon$4 carriers to enrich for likely decliners would exclude many potential candidates who are at high risk of decline based on other factors. Our predictive models can provide a more holistic view of an individual's future, which cannot be accurately forecasted by simple inclusion/exclusion criteria. As precision medicine progresses, clinical trials will need to reassess their recruitment strategies, as they currently depend on such coarse inclusion/exclusion criteria.

One potential limitation of our study is our use of observational cohorts, such as ADNI, for our training samples because it has been shown that placebo groups of clinical trials decline less on average compared to observational cohorts \cite{Berres2021-pz, Petersen2017-vh}. However, we did find that the early AD prediction model generalized well to a placebo arm of a past Phase 3 clinical trial (\autoref{fig:f1}). A second potential limitation is our lack of inclusion of markers of AD pathology, such as amyloid or tau, as input features. Approximately half of the early AD and a third of the presymptomatic participants had amyloid measurements. Due to the scarcity of AD pathology markers across the datasets in our study, we were unable to use such variables as input features in our models. While the inclusion of AD pathology markers has been shown to boost the performance of similar predictive models \cite{Cullen2021-tr}, a strength of our study is that our models rely on widely accessible biomarkers that clinical trials already depend on during the screening phase.

In summary, we presented prognostic models that can predict future cognitive decline from data that are routinely collected by clinical trials from a single time point. Our prognostic models can be applied to new clinical trials, alongside their inclusion/exclusion criteria, as an enrichment strategy to generate a cohort of individuals who are likely to decline and prevent the selection of individuals who are likely to remain stable. Active trials could use these models to assess endpoint imbalance from insufficient randomization of decliners and stable individuals among placebo and experimental arms and use the prognostic labels in randomization schemes. Lastly, we believe prognostic models such as ours can be used to guide clinical trials in their development, from the design stages to analysis phases, to increase their chances of meeting their endpoints.

%%%%%%%%%%%%%%%%%%%%%%%%%%%%%%%%%%%%%%%
% DECLARATIONS
%%%%%%%%%%%%%%%%%%%%%%%%%%%%%%%%%%%%%%%
\section{Conflicts of interest}
Dr Tam, Dr Laurent, and Dr Dansereau are employees of Perceiv Research Inc and hold stocks/stock options in Perceiv Research Inc. Dr Gauthier has nothing to disclose.

%%%%%%%%%%%%%%%%%%%%%%%%%%%%%%%%%%%%%%%
% ACKNOWLEDGEMENTS
%%%%%%%%%%%%%%%%%%%%%%%%%%%%%%%%%%%%%%%
\section{Acknowledgements}

Part of the data collection and sharing for this project was funded by the Alzheimer's Disease Neuroimaging Initiative (ADNI) (National Institutes of Health Grant U01 AG024904) and DOD ADNI (Department of Defense award number W81XWH-12-2-0012). ADNI is funded by the National Institute on Aging, the National Institute of Biomedical Imaging and Bioengineering, and through generous contributions from the following: AbbVie, Alzheimer’s Association; Alzheimer’s Drug Discovery Foundation; Araclon Biotech; BioClinica, Inc.; Biogen; Bristol-Myers Squibb Company; CereSpir, Inc.; Cogstate; Eisai Inc.; Elan Pharmaceuticals, Inc.; Eli Lilly and Company; EuroImmun; F. Hoffmann-La Roche Ltd and its affiliated company Genentech, Inc.; Fujirebio; GE Healthcare; IXICO Ltd.; Janssen Alzheimer Immunotherapy Research \& Development, LLC.; Johnson \& Johnson Pharmaceutical Research \& Development LLC.; Lumosity; Lundbeck; Merck \& Co., Inc.; Meso Scale Diagnostics, LLC.; NeuroRx Research; Neurotrack Technologies; Novartis Pharmaceuticals Corporation; Pfizer Inc.; Piramal Imaging; Servier; Takeda Pharmaceutical Company; and Transition Therapeutics. The Canadian Institutes of Health Research is providing funds to support ADNI clinical sites in Canada. Private sector contributions are facilitated by the Foundation for the National Institutes of Health (www.fnih.org). The grantee organization is the Northern California Institute for Research and Education, and the study is coordinated by the Alzheimer’s Therapeutic Research Institute at the University of Southern California. ADNI data are disseminated by the Laboratory for Neuro Imaging at the University of Southern California.

The NACC database is funded by NIA/NIH Grant U24 AG072122. NACC data are contributed by the NIA-funded ADRCs: P30 AG019610 (PI Eric Reiman, MD), P30 AG013846 (PI Neil Kowall, MD), P50 AG008702 (PI Scott Small, MD), P50 AG025688 (PI Allan Levey, MD, PhD), P50 AG047266 (PI Todd Golde, MD, PhD), P30 AG010133 (PI Andrew Saykin, PsyD), P50 AG005146 (PI Marilyn Albert, PhD), P50 AG005134 (PI Bradley Hyman, MD, PhD), P50 AG016574 (PI Ronald Petersen, MD, PhD), P50 AG005138 (PI Mary Sano, PhD), P30 AG008051 (PI Thomas Wisniewski, MD), P30 AG013854 (PI Robert Vassar, PhD), P30 AG008017 (PI Jeffrey Kaye, MD), P30 AG010161 (PI David Bennett, MD), P50 AG047366 (PI Victor Henderson, MD, MS), P30 AG010129 (PI Charles DeCarli, MD), P50 AG016573 (PI Frank LaFerla, PhD), P50 AG005131 (PI James Brewer, MD, PhD), P50 AG023501 (PI Bruce Miller, MD), P30 AG035982 (PI Russell Swerdlow, MD), P30 AG028383 (PI Linda Van Eldik, PhD), P30 AG053760 (PI Henry Paulson, MD, PhD), P30 AG010124 (PI John Trojanowski, MD, PhD), P50 AG005133 (PI Oscar Lopez, MD), P50 AG005142 (PI Helena Chui, MD), P30 AG012300 (PI Roger Rosenberg, MD), P30 AG049638 (PI Suzanne Craft, PhD), P50 AG005136 (PI Thomas Grabowski, MD), P50 AG033514 (PI Sanjay Asthana, MD, FRCP), P50 AG005681 (PI John Morris, MD), P50 AG047270 (PI Stephen Strittmatter, MD, PhD).

Data were provided in part by OASIS-3: Principal Investigators: T. Benzinger, D. Marcus, J. Morris; NIH P50 AG00561, P30 NS09857781, P01 AG026276, P01 AG003991, R01 AG043434, UL1 TR000448, R01 EB009352. AV-45 doses were provided by Avid Radiopharmaceuticals, a wholly owned subsidiary of Eli Lilly.

The PharmaCog project is funded through the European Community’s ‘Seventh Framework’ Programme (FP7/2007-2013) for an innovative scheme, the Innovative Medicines Initiative (IMI). IMI is a young and unique public-private partnership, founded in 2008 by the pharmaceutical industry (represented by the European Federation of Pharmaceutical Industries and Associations), EFPIA and the European Communities (represented by the European Commission).

%% Specify your .bib file name here, without the extension
\bibliography{CDRSB_refs}

%%%%%%%%%%%%%%%%%%%%%%%%%%%%%%%%%%%%%%%
% APPENDIX
%%%%%%%%%%%%%%%%%%%%%%%%%%%%%%%%%%%%%%%
\clearpage
\appendix
\renewcommand\thetable{S\arabic{table}}
\setcounter{table}{0}
\renewcommand\thefigure{S\arabic{figure}}
\setcounter{figure}{0}

\onecolumn
\section{Supplementary material}
%%% TABLE S1. Baseline characteristics per cohort
\begin{table}[H]
\centering
\begin{threeparttable}
\caption{Baseline characteristics of participants in each individual dataset}
\label{app:t1}
\begin{tabularx}{0.93\linewidth}{ 
@{}>{\smaller\raggedright}p{3.9cm} 
@{}>{\smaller\centering}p{1.5cm}
@{}>{\smaller\centering}p{1.5cm}
@{}>{\smaller\raggedright}p{0.45cm}
@{}>{\smaller\centering}p{1.5cm}
@{}>{\smaller\centering}p{1.5cm}
@{}>{\smaller\raggedright}p{0.45cm}
@{}>{\smaller\centering}p{1.5cm}
@{}>{\smaller\centering}p{1.5cm}
@{}>{\smaller\raggedright}p{0.45cm}
@{}>{\smaller\centering\arraybackslash}p{1.5cm}@{} }
\toprule
 & \multicolumn{5}{c}{\bf Early AD} & & \multicolumn{4}{c}{\bf Presymptomatic}\\
\cmidrule{2-6} \cmidrule{8-11}
 & \multicolumn{2}{c}{\bf Discovery} & & \multicolumn{2}{c}{\bf Validation} & & \multicolumn{2}{c}{\bf Discovery} & & \multicolumn{1}{c}{\bf Validation}\\
\cmidrule{2-3} \cmidrule{5-6} \cmidrule{8-9} \cmidrule{11-11}
                    & \bf ADNI      & \bf NACC      & & \bf Phase 3   & \bf PharmaCog & & \bf ADNI      & \bf NACC     & & \bf OASIS-3\\
\midrule
N                   & 858           & 293           & & 115           & 76            & & 237           & 391          & & 128\\
Age, mean (std)     & 73.3 (7.6)    & 72.7 (8.6)    & & 74.3 (7.9)    & 69.2 (7.5)    & & 74.3 (5.6)    & 66.2 (9.3)   & & 69.4 (5.4)\\
Female \%           & 41.6          & 49.1          & & 47.0          & 55.3          & & 51.9          & 68.3         & & 53.1\\
Education, mean (std) & 15.8 (2.8)  & 15.0 (3.4)    & & -             & 11.5 (4.5)    & & 16.3 (2.6)    & 16.2 (2.9)   & & 16.0 (2.4)\\
APOE $\epsilon$4 carriers \% & 54.0 & 51.5          & & 51.3          & 47.4          & & 27.8          & 32.7         & & 34.4\\
Baseline diagnosis & & & & & & & & & &\\
\hspace{2em}Cognitively unimpaired \% & 0.0 & 0.0   & & 0.0           & 0.0           & & 100.0         & 100.0        & & 100.0\\
\hspace{2em}Mild cognitive impairment \% & 80.7 & 57.3 & & 0.0        & 100.0         & & 0.0           & 0.0          & & 0.0\\
\hspace{2em}AD dementia \% & 19.3 & 42.7            & & 100.0         & 0.0           & & 0.0           & 0.0          & & 0.0\\
MMSE, mean (std)    & 26.8 (2.5)    & 24.6 (4.6)    & & 23.4 (1.7)    & 26.7 (1.8)    & & 29.1 (1.1)    & 29.2 (1.1)   & &28.9 (1.5)\\
CDR-SB, mean (std) & 2.04 (1.5)     & 3.11 (2.6)    & & 3.99 (1.6)    & 1.10 (0.8)    & & 0.03 (0.1)    & 0.10 (0.4)   & & 0.02 (0.1)\\
\bottomrule
\end{tabularx}
\begin{tablenotes}[flushleft]
\smaller
\item Abbreviations: AD, Alzheimer's disease; APOE, apolipoprotein E; CDR-SB, Clinical Dementia Rating Scale - Sum of boxes; MMSE, Mini-Mental State Examination.
\end{tablenotes}
\end{threeparttable}
\end{table}

%%% TABLE S2. Performance of EARLY AD models
\begin{table}[H]
\centering
\begin{threeparttable}
    \caption{Performance (validation AUC) of early AD models}
    \label{app:t2}
    \begin{tabularx}{0.43\linewidth}{ @{}l c@{} }
    \toprule
    \bf Model & \bf AUC (\%) \\
    \midrule
    Age, sex, APOE $\epsilon$4, MMSE, CDR-SB, MRI  & 78.0 $\pm$ 1.6\\
    Age, sex, APOE $\epsilon$4, MRI                & 76.3 $\pm$ 1.6\\
    Age, sex, APOE $\epsilon$4, MMSE, CDR-SB       & 75.2 $\pm$ 1.2\\
    Age, sex, APOE $\epsilon$4                     & 63.8 $\pm$ 1.2\\
    \bottomrule
    \end{tabularx}
\begin{tablenotes}[flushleft]
\smaller
\item Abbreviations: APOE, apolipoprotein E; AUC, area under the curve; CDR-SB, Clinical Dementia Rating Scale - Sum of boxes; MMSE, Mini-Mental State Examination; MRI, magnetic resonance imaging.
\end{tablenotes}
\end{threeparttable}
\end{table}

%%% TABLE S3. Performance of PRESYMPTOMATIC models
\begin{table}[H]
\centering
\begin{threeparttable}
    \caption{Performance (validation AUC) of presymptomatic models}
    \label{app:t3}
    \begin{tabularx}{0.53\linewidth}{ @{}l c@{} }
    \toprule
    \bf Model & \bf AUC (\%) \\
    \midrule
    Age, sex, APOE $\epsilon$4, education, FAQ, MMSE, CDR-SB, MRI  & 71.8 $\pm$ 1.2\\
    Age, sex, APOE $\epsilon$4, FAQ, MMSE, CDR-SB, MRI             & 71.8 $\pm$ 1.2\\
    Age, sex, APOE $\epsilon$4, MMSE, CDR-SB, MRI                  & 71.6 $\pm$ 1.1\\
    Age, sex, APOE $\epsilon$4, MRI                                & 71.3 $\pm$ 1.2\\
    Age, sex, APOE $\epsilon$4, education, MRI                     & 71.3 $\pm$ 1.3\\
    Age, sex, APOE $\epsilon$4, FAQ, MMSE, CDR-SB                  & 71.0 $\pm$ 1.5\\
    Age, sex, APOE $\epsilon$4, education, FAQ, MMSE, CDR-SB       & 70.5 $\pm$ 1.3\\
    Age, sex, APOE $\epsilon$4, MMSE, CDR-SB                       & 70.1 $\pm$ 1.5\\
    Age, sex, APOE $\epsilon$4                                     & 68.8 $\pm$ 1.7\\
    Age, sex, APOE $\epsilon$4, education                          & 68.3 $\pm$ 1.3\\
    \bottomrule
    \end{tabularx}
\begin{tablenotes}[flushleft]
\smaller
\item Abbreviations: APOE, apolipoprotein E; AUC, area under the curve; CDR-SB, Clinical Dementia Rating Scale - Sum of boxes; FAQ, Functional Activities Questionnaire; MMSE, Mini-Mental State Examination; MRI, magnetic resonance imaging.
\end{tablenotes}
\end{threeparttable}
\end{table}

\begin{figure}
    \centering
        \includegraphics[width=0.8\linewidth]{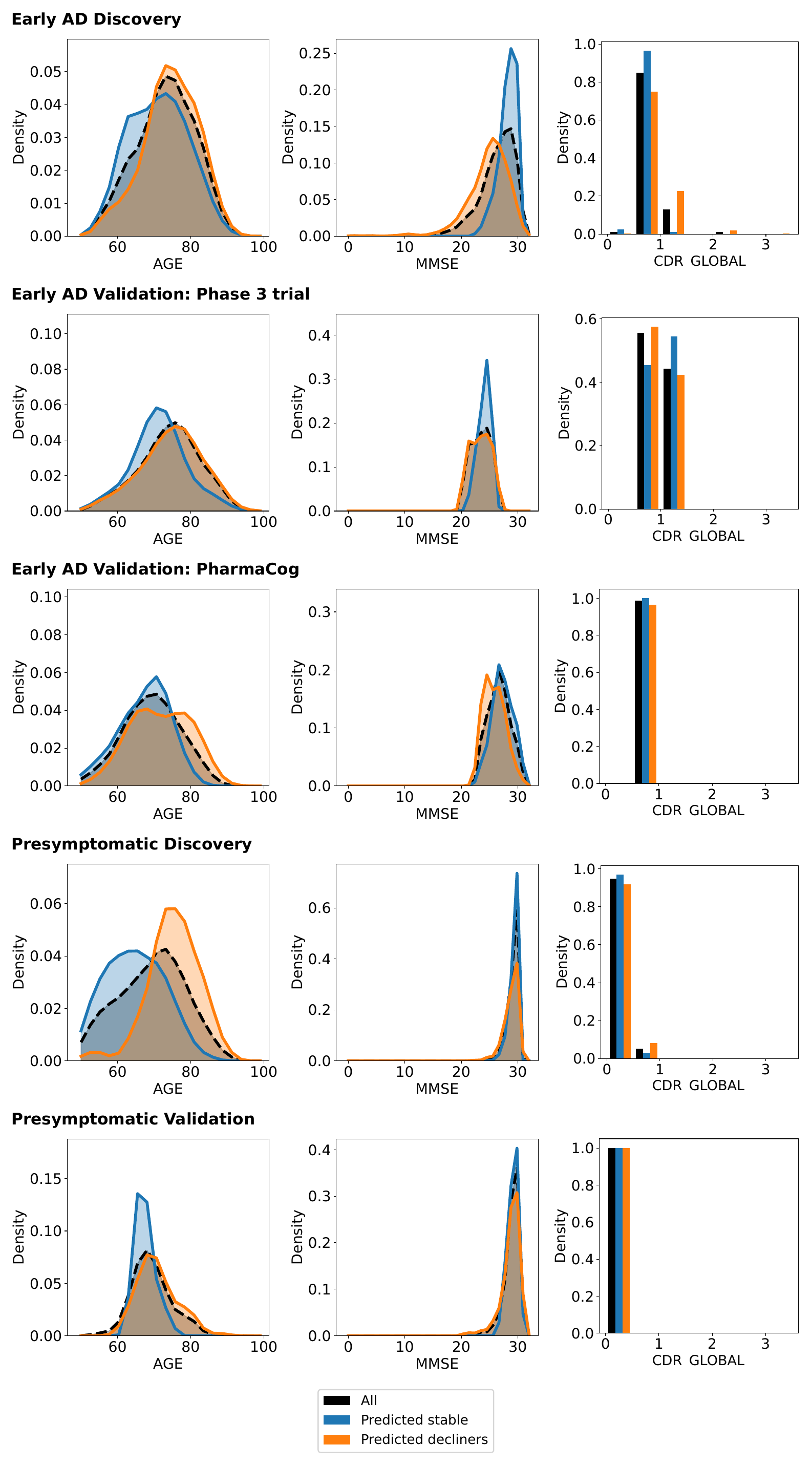}
        \captionsetup{width=0.8\columnwidth}
        \caption{Distributions of age, MMSE, and global CDR measured at baseline for the whole sample (black), the predicted stable (blue) and the predicted decliners (orange).}
        \label{app:f1}
\end{figure}

\begin{figure}
    \centering
        \includegraphics[width=0.8\linewidth]{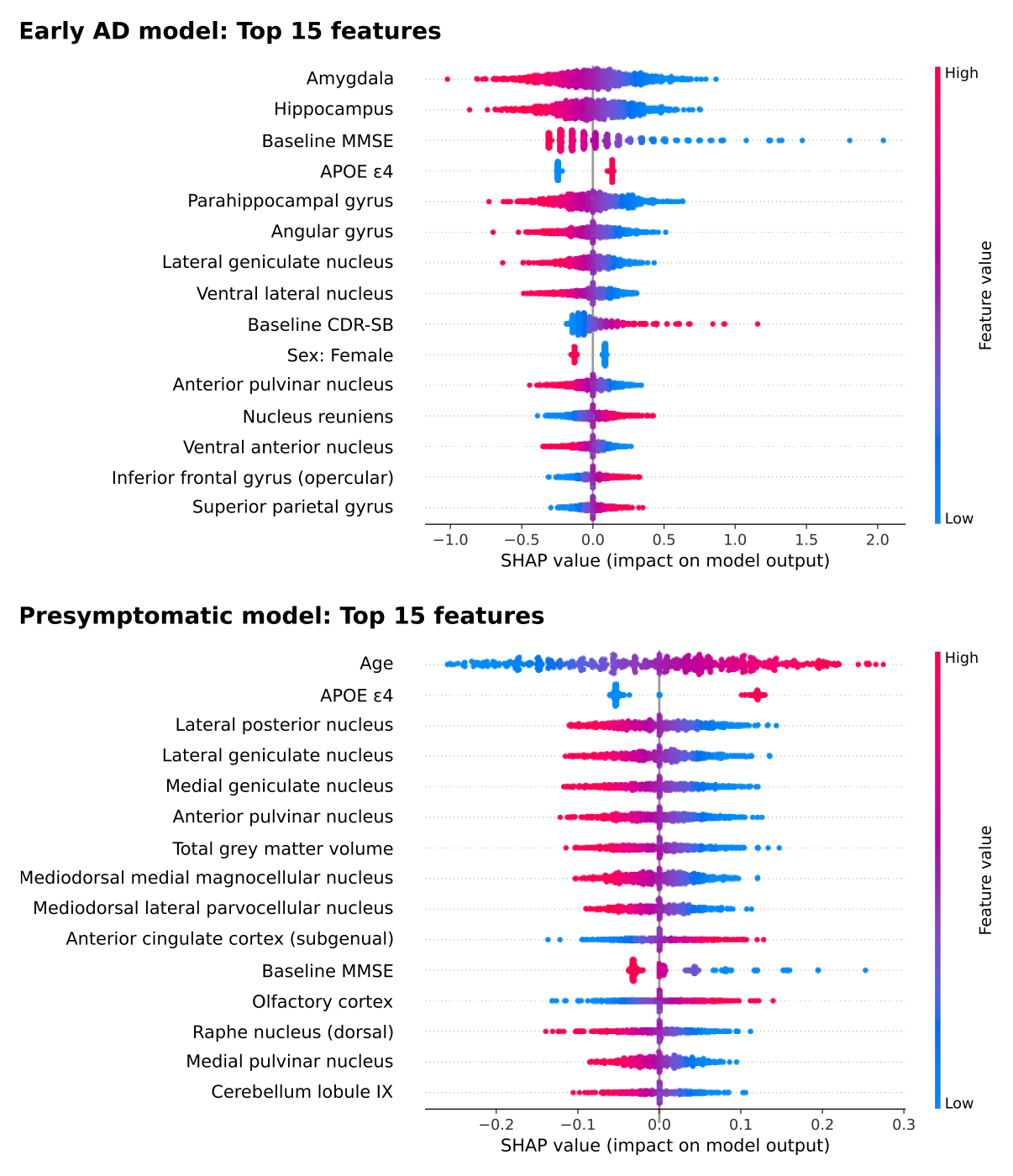}
        \captionsetup{width=0.8\columnwidth}
        \caption{Analysis of the importance of each feature for both the early AD (top) and presymptomatic (bottom) models. The x-axis contains SHapley Additive exPlanations (SHAP) values \cite{Lundberg2017-ve}, while the y-axis contains the top 15 most important features, in descending order. Each point in the plot denotes the feature of a sample from the discovery set, and its colour indicates the value of that feature: higher feature values are in red and lower feature values are in blue. SHAP values measure the impact of a feature on the output of the model: the higher the SHAP value, the more that feature is driving the model to classify that participant as a decliner, while the lower the SHAP value, the more that feature is driving the model to predict that the participant will remain stable. Looking at the presymptomatic model for instance, one can see that older individuals (i.e. with a higher age, thus in red) have higher SHAP values, while younger individuals (in blue) have lower SHAP values. This indicates that the older the individual, the more the model will be driven to predict that individual as a decliner.}
        \label{app:f2}
\end{figure}

\end{document}